# Some features of electromagnetic field of charged particle revolving about dielectric ball

**S R Arzumanyan, L Sh Grigoryan[1], H F Khachatryan and M L Grigoryan**
Institute of Applied Problems in Physics, 25 Hr. Nersessian Str., 0014 Yerevan, Armenia

E-mail: levonshg@mail.ru

**Abstract.** A relativistic electron uniformly rotating along an equatorial orbit around a dielectric ball may generate Cherenkov radiation tens of times more intense as that in case of revolution of a particle in a continuous, infinite and transparent medium. The root-mean-square values of electric and magnetic field strengths of particle are practically not localized in the central part of the equatorial plane of ball and close to the poles of ball.

## 1. Introduction and background

The presence of matter may essentially influence the characteristics of high energy electromagnetic processes giving rise to the production of Cherenkov radiation (CR), transition radiation etc. [1-9]. The operation of a number of devices assigned to production of electromagnetic radiation is based on the interaction of relativistic electrons with matter (see e.g. [8,9]).

The effects of interest arise in stratified media of different configurations (refer, e.g., [10-12] and references therein). The interfaces of media are widely used to control the radiation flow emitted by various systems. In the present work the case of one-layer medium of spherical configuration has been investigated.

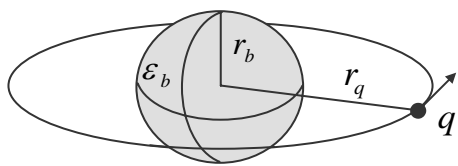

**Figure 1.** A relativistic particle rotating about a dielectric ball in it's equatorial plane.

Now consider a relativistic particle uniformly rotating in the magnetic field in vacuum about a dielectric ball in its equatorial plane. In spherical coordinates $r, \theta, \varphi$ with origin in the center of ball, the permittivity $\varepsilon(r)$ of the substance is a step function of radial coordinate:

$$\varepsilon(r) = \varepsilon_b \quad \text{for } r \leq r_b \quad \text{and} \quad \varepsilon(r) = 1 \text{ for } r > r_b, \qquad (1)$$

---

[1] To whom any correspondence should be addressed.

where $\varepsilon_b = \varepsilon_b' + i\varepsilon_b''$ is the complex valued permittivity of ball with radius $r_b$. The magnetic permeability is taken to be 1. The current density may be written as

$$\vec{j}(\vec{r},t) = \frac{q\mathrm{v}\vec{e}_\varphi}{r_q^2}\delta(r - r_q)\delta(\theta - \pi/2)\delta(\varphi - \omega_q t), \qquad (2)$$

where $q$, $\mathrm{v} = r_q\omega_q$ and $\omega_q$ are the charge, linear velocity and cyclic frequency of particle rotation, and $r_q$ is the radius of particle orbit.

The rotation of particle entails radiation at some discrete frequencies (harmonics) $\omega_k = k\omega_q$ with $k = 1;2;3...$ We assumed that an exterior force would make up for the braking of particle due to the radiation, by forcing the particle to uniformly rotate about the ball.

At large distances from the ball the radiation intensity $I_k$ after averaging over the period $T = 2\pi/\omega_q$ of revolution is determined by the expression [13]

$$I_k = \frac{c}{2\pi}\lim_{r\to\infty}r^2\int\left|\mathrm{rot}\vec{A}_k(\vec{r})\right|^2 d\Omega, \qquad (3)$$

where $\Omega$ is the solid angle, and $\vec{A}_k(\vec{r})$ is the Fourier component of vector-potential of the electromagnetic field that satisfies the equation

$$(\Delta + \frac{\omega_k^2}{c^2}\varepsilon)\vec{A}_k(\vec{r}) - \frac{1}{\varepsilon}(\vec{\nabla}\varepsilon)\mathrm{div}\vec{A}_k(\vec{r}) = -\frac{4\pi}{c}\vec{j}_k(\vec{r}). \qquad (4)$$

It is convenient to introduce a non-dimensional quantity

$$TI_k/\hbar\omega_k \equiv n_k, \qquad (5)$$

where $TI_k$ is the energy emitted at $\omega_k$ frequency during one period of particle revolution, and $\hbar\omega_k$ is the energy of quantum of corresponding electromagnetic wave. So, $n_k$ is the "number of electromagnetic field quanta" emitted during one revolution period of particle.

The dependence of $n_k$ on $\mathrm{v}$, $r_b/r_q$ and $\varepsilon_b$ was studied in [14]. It was shown that at some harmonics, in case of weak absorption of radiation in the ball material ($\varepsilon_b'' \ll \varepsilon_b'$), the particle may generate $n_k \sim 1$ radiation field quanta exceeding in several dozens of times those generated by particle rotating in a continuous, infinite and transparent medium having the same permittivity as the ball material ($\varepsilon = \varepsilon_b'$). In the present paper we have studied an angular distribution of electric and magnetic field strengths of particle inside the ball and on its surface.

## 2. Basic formulae
Due to the azimuthal symmetry of system one may write the following Fourier expansions

$$\vec{E}(r,\theta,\varphi,t) = \sum_{s=-\infty}^{\infty}[E_{sr}(r,\theta)\vec{e}_r + E_{s\theta}(r,\theta)\vec{e}_\theta + E_{s\varphi}(r,\theta)\vec{e}_\varphi]\exp[is(\varphi - \omega_q t)]$$

$$\vec{H}(r,\theta,\varphi,t) = \sum_{s=-\infty}^{\infty}[H_{sr}(r,\theta)\vec{e}_r + H_{s\theta}(r,\theta)\vec{e}_\theta + H_{s\varphi}(r,\theta)\vec{e}_\varphi]\exp[is(\varphi - \omega_q t)] \qquad (6)$$

for the strengths of electric and magnetic fields of particle ($\vec{e}_r$, $\vec{e}_\theta$ and $\vec{e}_\varphi$ are the unit vectors of the spherical system of coordinates). For this reason the root-mean-square values of electromagnetic field strengths may be written in the following form

$$\frac{1}{T}\int_0^T \vec{F}^2(\vec{r},t)dt = F_0^2(r,\theta) + \sum_{k=1}^\infty F_k^2(r,\theta), \qquad (7)$$

where

$$F_0^2 = |F_{0r}|^2 + |F_{0\theta}|^2 + |F_{0\varphi}|^2, \quad F_k^2 = 2(|F_{kr}|^2 + |F_{k\theta}|^2 + |F_{k\varphi}|^2), \qquad k=1,2,3,\ldots \quad (8)$$

and $F = E, H$. The problem is reduced to the calculation of $E_{kr}, E_{k\theta}, E_{k\varphi}$ and $H_{kr}, H_{k\theta}, H_{k\varphi}$.

In our calculations we took advantage of the method for obtaining appropriate exact solutions of equation (4) proposed in [15]. Final expressions for $F_{kr}, F_{k\theta}, F_{k\varphi}$ inside a ball are given below:

$$E_{kr} = \frac{ic}{\omega_k \varepsilon_b r}\sum_{l=k}^\infty p_{lk}^{(2)}(r)Y_{lk}(\theta,0), \qquad E_{k\theta} = \sum_{l=k}^\infty [p_{lk}^{(1)}(r)g_{lk}(\theta)\frac{ir\omega_k}{cl(l+1)} + p_{lk}^{(4)}(r)d_{lk}(\theta)],$$

$$E_{k\varphi} = \sum_{l=k}^\infty [p_{lk}^{(4)}(r)g_{lk}(\theta) - p_{lk}^{(1)}(r)d_{lk}(\theta)\frac{ir\omega_k}{cl(l+1)}] \qquad (9)$$

for an electric field and

$$H_{kr} = \sum_{l=k}^\infty p_{lk}^{(1)}(r)Y_{lk}(\theta,0), \quad H_{k\theta} = \sum_{l=k}^\infty [p_{lk}^{(2)}(r)\frac{g_{lk}(\theta)}{l(l+1)} + p_{lk}^{(3)}(r)d_{lk}(\theta)],$$

$$H_{k\varphi} = \sum_{l=k}^\infty [p_{lk}^{(3)}(r)g_{lk}(\theta) - p_{lk}^{(2)}(r)\frac{d_{lk}(\theta)}{l(l+1)}] \qquad (10)$$

for a magnetic field. Here we introduce the following notations

$$p_{lk}^{(1)}(r) = 4\pi q v d_{lk}(\pi/2)\underline{h_l}(\sigma u)\frac{\underline{j_l}(\tau u_b)}{cr_b^2 \tau}, \qquad p_{lk}^{(2)}(r) = \underline{j_l}(\tau u_b)\frac{4\pi q v u_b c_{lk} g_{lk}(\pi/2)}{cr_b^2(2l+1)},$$

$$p_{lk}^{(3)}(r) = [(l+1)\underline{j_{l-1}}(\tau u_b) - l\underline{j_{l+1}}(\tau u_b)]\frac{4\pi q v u_b d_{lk}(\pi/2)\underline{h_l}(\sigma u)}{cr_b^2 l(l+1)(2l+1)}, \qquad (11)$$

$$p_{lk}^{(4)}(r) = [(l+1)\underline{j_{l-1}}(\tau u_b) - l\underline{j_{l+1}}(\tau u_b)]\frac{4\pi i q v \omega_k c_{lk} g_{lk}(\pi/2)}{r_b c^2 l(l+1)(2l+1)^2},$$

where $h_l(y) = j_l(y) + in_l(y)$ ($j_l$ and $n_l$ are spherical Bessel and Neumann functions respectively), $Y_{lm}(\theta,\varphi)$ is the spherical function,

$$\sigma = \frac{r_q}{r_b}, \quad u = \frac{\omega_k r_b}{c}, \quad \tau = r/r_b \leq 1, \quad u_b = \frac{\omega_k r_b}{c}\sqrt{\varepsilon_b},$$

$$g_{lk}(\theta) = ikY_{lk}(\theta,0), \qquad d_{lk}(\theta) = \frac{\partial Y_{lk}(\theta,0)}{\partial \theta}, \qquad (12)$$

$$\underline{f_l}(\tau) = f_l(\tau)/[u_b j_{l+1}(u_b)h_l(u) - j_l(u_b)uh_{l+1}(u)],$$

and

$$c_{lk} = (l+1)h_{\underline{l-1}}(\sigma u) - lh_{\underline{l+1}}(\sigma u) +$$

$$+ (\varepsilon_b - 1)[h_{\underline{l-1}}(u) + h_{\underline{l+1}}(u)][h_{\underline{l-1}}(\sigma u) + h_{\underline{l+1}}(\sigma u)]\frac{l(l+1)u_b j_l(u_b)}{l\gamma_{l-1}^l + (l+1)\gamma_{l+1}^l}. \qquad (13)$$

In the last expression

$$\gamma_v^l = \frac{\varepsilon_b j_v(u_b)uh_l(u) - u_b j_l(u_b)h_v(u)}{j_v(u_b)uh_l(u) - u_b j_l(u_b)h_v(u)}. \qquad (14)$$

## 3. Numerical results and discussion

Let us consider an electron uniformly revolving about a dielectric ball. The electromagnetic field of electron is determined by its energy $E_q$ and parameters $r_q$, $r_b$ and $\varepsilon_b$. It is also convenient to introduce a dimensionless parameter

$$\nu = \text{v} r_b / c r_q \qquad (15)$$

and assume for certainty in what follows that $\varepsilon_b = 3.78$ (fused quartz in the range of microwaves), $r_b = 36.2 mm$, $r_q \cong 36.9 mm$, $k = 8$ and $E_q = 2MeV$.

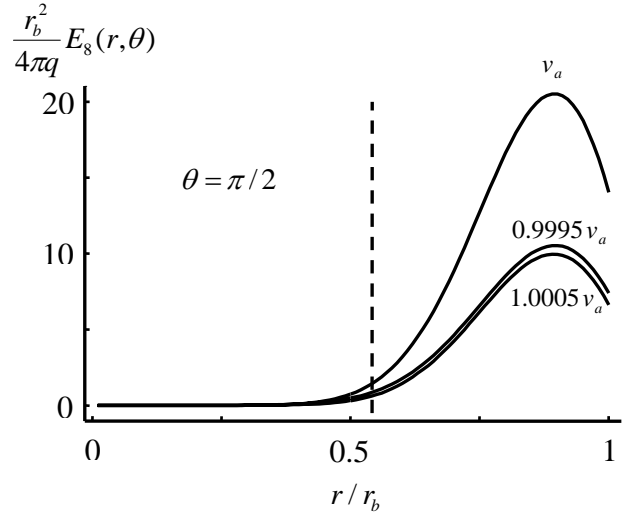

**Figure 2.** $r$-dependence of the root-mean-square strength $E_k(r,\theta)$ of the $k$-th harmonic of electric field inside the ball at $\theta = \pi/2$ (the equatorial plane of the ball) for $2MeV$ electron. The values of parameter (15) are shown beside the curves, $\nu_a = 1.2507 Ghz$ [14], $\varepsilon_b = 3.78$, $k = 8$.

In figure 2 the root-mean-square strength $E_k(r,\theta)$ of the electric field inside a dielectric ball at the equatorial plane ($\theta = \pi/2$) is plotted versus the ratio $r/r_b$. Here we give three curves of $r$-dependence of $E_k$ for $\nu = \nu_a$; $0.9995\nu_a$ and $1.0005\nu_a$. The material of ball is assumed to be transparent ($\varepsilon_b'' = 0$). It is seen that large values of $E_k(r)$ are possible on the curves with $\nu \cong \nu_a$. The comparison of curves with $\nu/\nu_a = 1$; $0.9995$ and $1.0005$ shows how rapidly the function $E_k(r)$ tends to the maximum when $\nu \to \nu_a$. A slight departure of $\nu$ in either direction from $\nu_a$ would suffice for the point describing the state of system to notably fall. For this reason one may term $\nu_a$ as *the resonance value of $\nu$ parameter*.

As it is seen from the plots in figure 2, the function $E_k(r)$ steeply increases rightwards of the vertical dashed line. Corresponding to this line is the value

$$r/r_b = 0.54 \cong \cos\theta_C, \quad \text{where} \quad \theta_C = \arccos(c/\text{v}\sqrt{\varepsilon_b}) \qquad (16)$$

is the Cherenkov radiation emission angle for an electron at rectilinear motion in a continuous medium with $\varepsilon = \varepsilon_b$. This fact testifies to the relation of localization of the electromagnetic field in the range of $\cos\theta_C \leq r/r_b \leq 1$ to the generation of CR inside the dielectric ball.

In figure 3 the electric field strength $E_k$ is plotted versus $\theta$ at the surface of the ball ($r = r_b$). As in figure 2 here we give three curves of $\theta$- dependence of $E_k$ for $v = v_a$; $0.9995v_a$ and $1.0005v_a$. It is seen again that large values of $E_k(\theta)$ are possible on the curves with $v \cong v_a$.

As it is seen from the plots in figure 3, the function $E_k(\theta)$ increases sufficiently fast rightwards of the vertical dashed line. Corresponding to this line is the value
$$\theta = \arccos(\sin^2\theta_C) \cong 44.5^o. \quad (17)$$

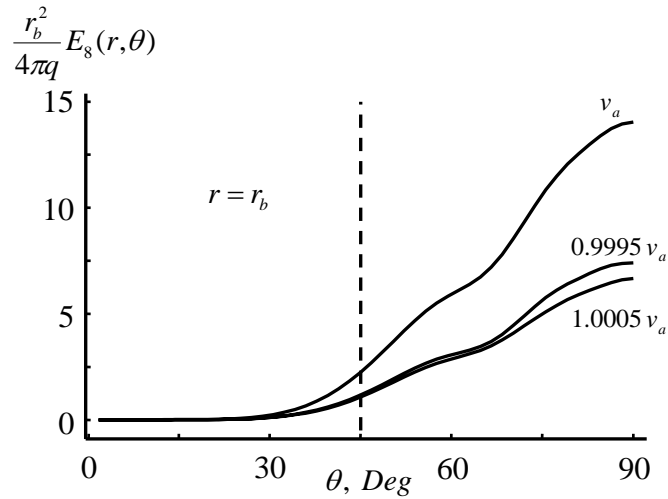

**Figure 3.** The root-mean-square strength $E_8(r,\theta)$ of the 8-th harmonic of electric field on the surface of ball ($r = r_b$) as the function of polar angle $\theta$. The notations and values of parameters are the same as those in figure 2.

This fact again testifies to the relation of localization of the electromagnetic field in the range of
$$\arccos(\sin^2\theta_C) \leq \theta \leq 90° \quad (18)$$
to the generation of CR inside the dielectric ball.

The results of our numerical calculations indicate that close to the maxima in figures 2,3 the value of $E_k$ are tens of times as large as that in the case of revolution of a particle in a continuous, infinite and transparent medium having the same permittivity as the ball material. The situation with the root-mean-square strength $H_k(r,\theta)$ of the magnetic field is simalar. Sure, the electromagnetic oscillations of CR induced by the particle along the trajectory on the whole are partially locked inside the ball by superimposing each other mostly in the destructive way. However, in some specific cases ($v \cong v_a$ in figures 2,3) the oscillations may superimpose nondestructively, as a result of which the electromagnetic field of particle inside the ball is amplified.

The maxima of functions $E_k(r,\theta)$ and $H_k(r,\theta)$ at resonance are highly sensitive to the values of permittivity of the ball and its radius. This fact may be used in applications.

**References**

[1] Jelley J V 1958 *Cherenkov Radiation and its Applications* (London: Pergamon Press)
[2] Bolotovskii B M 1961 *Physics-Uspekhi* 75 295
[3] Zrelov V I 1968 *Vavilov-Cherenkov Radiation (and its Applications in High Energy Physics)* (Moscow: Atomizdat) (in Russian)
[4] Garibian G M and Yang C 1983 *X-Ray Transition Radiation* (Yerevan: AN Arm. SSR Press) (in Russian)
[5] Ginzburg V L and Tsytovich V N 1990 *Transition Radiation and Transition Scattering* (Bristol: Adam Hilger)
[6] Frank I M 1988 *Vavilov-Cherenkov Radiation. Theoretical Aspects* (Moscow: Nauka) (in Russian)
[7] Berman B L and Datz S 1985 in: *Coherent Radiation Sources* eds A W Saenz and H Uberall (Berlin: Springer-Verlag)
[8] Rullhusen P, Artru X and Dhez P 1998 *Novel Radiation Sources Using Relativistic Electrons* (Singapore: World Scientific)
[9] Potylitsyn A P 2009 *Radiation from Electrons in Periodic Structures* (Tomsk: NTL Press) (in Russian)
[10] Saharian A A and Kotanjyan A S 2009 *J. Phys. A: Math. Theor.* **42** 135402
[11] Arzumanyan S R, Grigoryan L Sh, Khachatryan H F and Grigoryan M L 2008 *Nucl. Instr. and Meth.* B **266** 3715–3720
[12] Arzumanyan S R, Grigoryan L Sh, Khachatryan H F, Kotanjyan A S and Saharian A A *2008 Nucl. Instr. and Meth.* B **266** 3703–3707
[13] Landau L D and Lifshitz E M 1980 *The Classical Theory of Field* (Oxford: Pergamon Press)
[14] Grigoryan L Sh, Khachatryan H F, Arzumanyan S R and Grigoryan M L 2006 *Nucl. Instr. and Meth.* B **252** 50–56
[15] Arzumanyan S R, Grigoryan L Sh, Saharian A A, Kotanjian Kh V *Izv. Nats. Akad. Nauk Arm Fiz.* (Engl. Transl.: *J. Contemp. Phys.*) **30** 106-113

**Acknowledgements**

L Sh Grigoryan is thankful to K V Lekomtsev for useful discussion. This work was supported in part by the Grant No. 077 from Ministry of Education and Science of RA.